# Oxygen Vacancies at Dislocation Core Modulate Plasticity in Strontium Titanate

Min-Chul Kang[1], Chunxu Yan[2], Alexander Frisch[3], Xufei Fang[3*], Liming Xiong[2*], Lin Zhou[1,4,5*]

[1] Ames National Laboratory, 50011, Ames, IA, United States

[2] Department of Mechanical and Aerospace Engineering, North Carolina State University, 27606, Raleigh, NC, United States

[3] Institute for Applied Materials, Karlsruhe Institute of Technology, 76131, Karlsruhe, Germany

[4] Department of Materials Science and Engineering, Iowa State University, 50011, Ames, IA, United States

[5] Department of Materials Science and Engineering, University of Virginia, 22903, Charlottesville, VA, United States

*Corresponding authors: linzhou@virginia.edu, xufei.fang@kit.edu, lxiong3@ncsu.edu

## Abstract

Dislocation core chemistry in oxides critically influences mechanical behavior and functionality; yet the evolution of core chemistry during the dislocation motion in them has not been directly observed. Here, using $SrTiO_3$ as a model material, we combine aberration-corrected scanning transmission electron microscopy and electron energy-loss spectroscopy with atomic-level molecular dynamics (MD) simulations to correlate the $\langle 110\rangle\{\bar{1}10\}$ dislocation core structure, oxygen vacancy density, charge state, and mobility with each other. We find that the mechanically induced dislocation loops exhibit dissociated cores, whose oxygen vacancy density depends on the gliding distance: short loops are Ti-reduced and oxygen-deficient at the edge dislocation core, whereas longer loops remain close to stoichiometry in both the edge and screw components. MD simulations reveal that kink-assisted edge dislocation glide in $SrTiO_3$ leaves oxygen-deficient trails behind, modulating the oxygen content inside the edge core. These results demonstrate that oxygen-vacancy evolution at the dislocation core intrinsically couples with plasticity in ionic crystals, suggesting a mechanism for oxygen vacancy-dependent dislocation mobility in plastically deformed oxides.

## Keywords

Dislocations; Oxygen deficiency; Strontium Titanate; STEM-EELS; Atomistic Simulations

## Graphical abstract

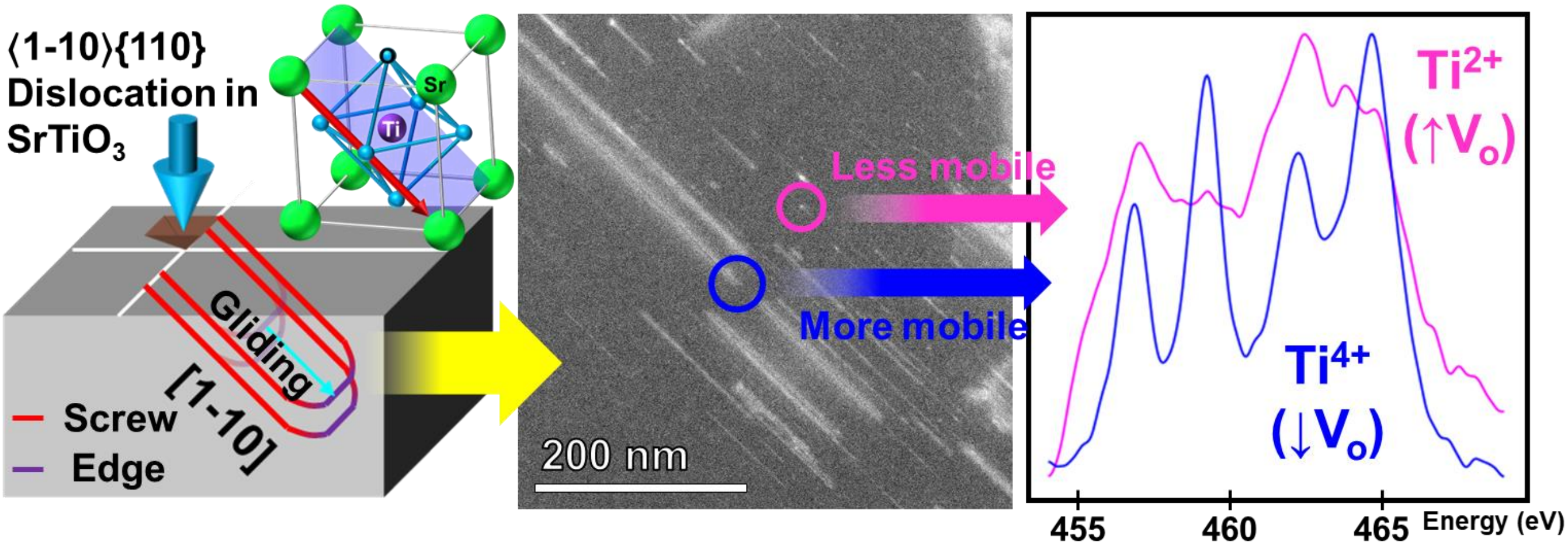

## 1. Introduction

Dislocations in complex oxides not only control their plastic deformation behavior but also alter their functionality. For instance, in strontium titanate ($SrTiO_3$), a model perovskite oxide for versatile fundamental studies of the electronic and optical properties, the chemistry of dislocation cores can reconfigure the surrounding electronic structure [1], creating band bending [2], defect-derived states [3], and pathways for carrier accumulation or depletion [4]. Such dislocation cores, often accompanied by local redox variations and oxygen vacancies, can act as quasi-one-dimensional conduction channels [4,5], scattering centers [6,7], or even localized superconducting [8–13] or photoactive sites [1,14,15]. Understanding and controlling the chemistry of dislocations is therefore pivotal for advanced design of functional oxides, where defect dynamics and electronic/ionic transport are intimately coupled [3,12,16].

$SrTiO_3$ was shown nearly two decades ago to be plastically deformable at room temperature, with a bulk yield strength of ~120 MPa [17,18]. Later, the lattice friction stress of edge dislocation in $SrTiO_3$ at room temperature was experimentally evaluated to be 89 MPa [19]. Tuning the stoichiometry (Sr/Ti ratio in the starting powder for growing the single crystals) and the oxygen vacancy concentration in $SrTiO_3$ can strongly modify its room-temperature plasticity [20–22]. Simulations have indicated that dislocation mobility is strongly dependent on defect chemistry: including oxygen vacancy and the resulting core electrostatics often described as space-charge effects in ionic models [23–25]. $SrTiO_3$ is no exception. More perovskite oxides, including $KNbO_3$ and $KTaO_3$ [26,27], have been recently reported to exhibit dislocation-governed, room-temperature bulk plasticity. A pertinent question arises: how does the chemistry of a dislocation core (e.g., oxygen deficiency and local redox) differ from that of a bulk, and why does it evolve during dislocation motion? This highlights one key distinction between metals with metallic bonding and oxides with ionic bonding, as the past discussions on dislocation mobility in oxides are dominated by simply referring to the mechanisms in metals for analogy. More importantly, the chemistry of dislocations in oxides is responsible for tuning the versatile physical properties. These aspects position oxygen vacancy at dislocation cores of dislocations as a critical parameter for linking dislocation mechanics with electronic/ionic transport properties. Nevertheless, the coupling between dislocation mobility and dislocation core chemistry has largely relied on simplified theories, lacking direct experimental evidence to correlate dislocation core chemistry with the behavior of individual dislocation line or loop. This leads to barriers when a validation of material models or a manipulation of the dislocation chemistry for tuning the performance of oxides becomes desired.

Here, we correlate the dislocation core oxygen vacancy and its mobility in $SrTiO_3$ through high-fidelity experiments. Using aberration-corrected scanning transmission electron microscopy (STEM) coupled with electron energy loss spectroscopy (EELS), we map Ti valence variations as an indicator of local reduction and oxygen deficiency near the $\langle 110\rangle\{\bar{1}10\}$ dislocations in plastically deformed $SrTiO_3$ at room temperature [28]. In parallel, atomic-level molecular dynamics (MD) simulations reveal how kink-mediated glide modulates the local oxygen deficiency, which is then correlated with the dislocation mobility in turn.

## 2. Experimental methods

Dislocations were introduced via Vickers indentation on the (010) surface in nominally undoped $SrTiO_3$ single crystal (Alineason Materials Technology GmbH, Frankfurt am Main, Germany) at room temperature. The target indentation load of 0.025 kgf was reached within 7 s upon surface contact and was held for 10 s before unloading. A total of 16 indents were placed with a distance of 30 μm. TEM sample was prepared using focused ion beam (FIB, FEI Helios NanoLab G3) in the nearby indented region (**Figure 1d**). S/TEM imaging was performed at 200 kV on a Titan Themis. High-angle annular-dark-field (HAADF) STEM images were acquired with a scattering angle of 74 to 200 mrad, and medium-angle annular dark field (MAADF) images were taken with a scattering angle range from 23 to 140 mrad. EELS spectra were collected with a current of 40 pA, covering an energy range of 446 to 548.4 eV with 2048 channels. To identify the EELS spectrum with reference peaks [29,30], non-negative least square (NNLS) was employed to ensure that all component weights remained non-negative, preventing unphysical negative contributions in the spectral decomposition [31,32]. Principal component analysis (PCA) was performed using singular value decomposition (SVD) in Python sklearn.decomposition module with two components [33]. The details about PCA are described in Supplementary Materials Part 1 (**Figure S1**).

MD simulations were performed in LAMMPS using a rigid-ion potential. [34] The simulation cell ($276 \times 276 \times 17$ nm$^3$) under consideration here contains ⟨110⟩{1-10} dislocations which are initially built inaccording to a procedure in [25]. In detail, firstly, a specific ribbon of unit cells is removed from the atomistic system. This extraction exposes an upper cleavage surface with an $SrTiO_3$ stoichiometry, which inherently possesses an excess positive charge. Subsequently, a mid-plane cut is executed by deleting two adjacent atomic layers: an SrO plane and a $TiO_2$ plane. This specific removal leaves the opposing bottom surface terminating completely at an oxygen ion plane, resulting in a net negative charge. These steps successfully lead to a non-stoichiometric interfaces and initial electrostatic conditions for the dislocation prior to its relaxation, which is then equilibrated for 0.3 ns at 300 K. Local charge density was obtained by Gaussian smoothing of atomic charges, and an effective Ti charge was defined from deviations in O coordination and linearly scaled to the +1~+4 range for comparison with experiment. See Supplementary Materials Part 2 **(Figure S4**) for full details.

## 3. Results and Discussion

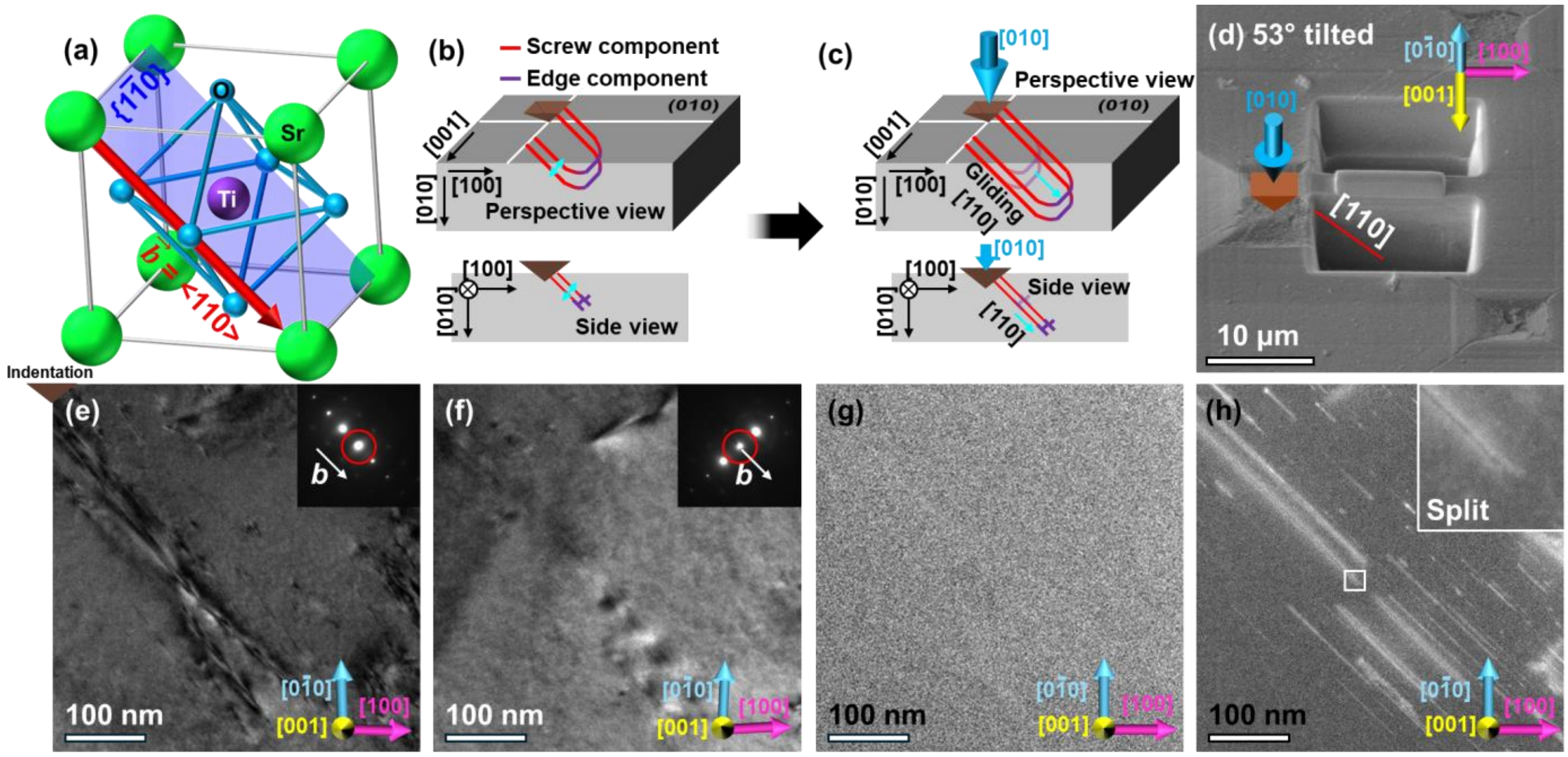

**Figure 1. Indentation and overall structure of a ⟨110⟩{1̄10} dislocation loop.** (a) Schematic diagram of ⟨110⟩{1̄10} slip system of $SrTiO_3$. (b) ⟨110⟩{1̄10} dislocation loop introduced by [010] direction indentation. (c) Gliding motion of dissociated ⟨110⟩{1̄10} dislocation loop. (d) SEM image of 0.025 kgf Vickers indents and the FIB lamella lift-out position. (e-h) Structures of dissociated ⟨110⟩{1̄10} dislocation observed by (e) [110] axis 3° tilted [001] zone axis BF-TEM image, (f) [1-10] axis 3° tilted [001] zone axis BF-TEM image, (g) HAADF STEM image, (h) MAADF STEM image, respectively.

In $SrTiO_3$, room-temperature mechanical deformation generates dislocations belonging to the ⟨110⟩{1̄10} slip systems (**Figure 1a**) [18,35,36]. This schematically depicts the atomic structure of the slip system and the ⟨110⟩ Burgers vector for reference. Under indentation, dislocation half loops with a ⟨110⟩ Burgers vector and on the {110} planes are emitted. The line sense of the dislocation will be perpendicular to the Burgers vector on some parts of the half loop near the surface and parallel to it on other part away from the surface, as the Burgers vector remains the same on the entire half loop. Therefore, both edge and screw components are formed (**Figure 1b**) [37]. These dislocations are further dissociated at room temperature with a stacking fault of a few nm between the partial dislocations [38]. When subjected to higher stress, as illustrated in **Figure 1c**, they further glide while maintaining their dissociated configuration. A TEM lamella adjacent to the [010] direction Vickers indentation (**Figure 1d**) was lifted out, and the gliding behavior of the $\langle 110\rangle\{\bar{1}10\}$ dislocation loop along the [110] direction is investigated by the bright field transmission electron microscopy (BF-TEM) images in **Figure 1e,f** (Detailed experimental processes are provided in the method section). **Figure 1e,f** were obtained by tilting the specimen by approximately 3° from the [001] zone axis to obtain the $\vec{g}_{110}$ and $\vec{g}_{1\bar{1}0}$ directions, respectively. Under the two-beam BF-TEM condition, the dislocation contrast visible in **Figure 1e**, obtained by $\vec{g}_{110}$ condition is not clearly visible in **Figure 1f**, obtained by $\vec{g}_{1\bar{1}0}$ condition. Considering the $\vec{g} \cdot \vec{b} = 0$invisibility criterion, this contrast change indicates that the dislocation has a [110] Burgers vector. In addition, since the dislocation line direction is also along [110], satisfying $\vec{u} \parallel \vec{b}$, the dislocation component mainly observed in the low-magnification BF-TEM image is a screw component.

In contrast, no distinct features were detected in the HAADF image in **Figure 1g**. The dislocation morphology is better visualized using medium angle annular dark field (MAADF) STEM imaging, which is more sensitive to lattice strain (**Figure 1h**). As schematically illustrated in **Figure 1c**, the inset of **Figure 1h** likewise resolves two partials separated by a few unit-cells, indicating that glide proceeds in the dissociated state. Among these dislocations, some exhibit relatively long gliding distances of several hundred nanometers, whereas the rest remain nearly point-like, corresponding to different degrees of mobility. In the following, we classify these as “long” and “short” loops, respectively, based on their observed glide distances. Short loops show glide distances below ~5 nm and appear nearly point-like in MAADF, whereas long loops glide beyond ~30–40 nm (typically hundreds of nm and occasionally exceeding ~1 μm). To correlate the oxygen vacancy of a dislocation core with its mobility, we acquired STEM-EELS spectrum at a relatively higher magnification near the dislocation core. EELS is a powerful spectroscopy technique for identifying the valence states because it detects the energy losses of inelastically scattered electrons, which correspond to the specific electronic transitions in atoms. The fine structure of core-loss edges, such as the Ti L-edge or O K-edge, reveals shifts in peak positions and spectral shapes are sensitive to oxidation states (e.g., $Ti^{2+}$, $Ti^{3+}$, $Ti^{4+}$), allowing a direct determination of the valence states at the atomic scale. In $SrTiO_3$, Sr is effectively fixed at +2; it participates via Sr vacancies rather than by changing valence. In contrast, Ti, as a transition metal, exhibits variable oxidation states depending on the oxygen vacancy environment [39–41]. This property enables Ti L-edge EELS analysis to serve as a reliable probe detecting local redox chemistry (Ti valence) and associated oxygen vacancy-related chemistry near dislocations,

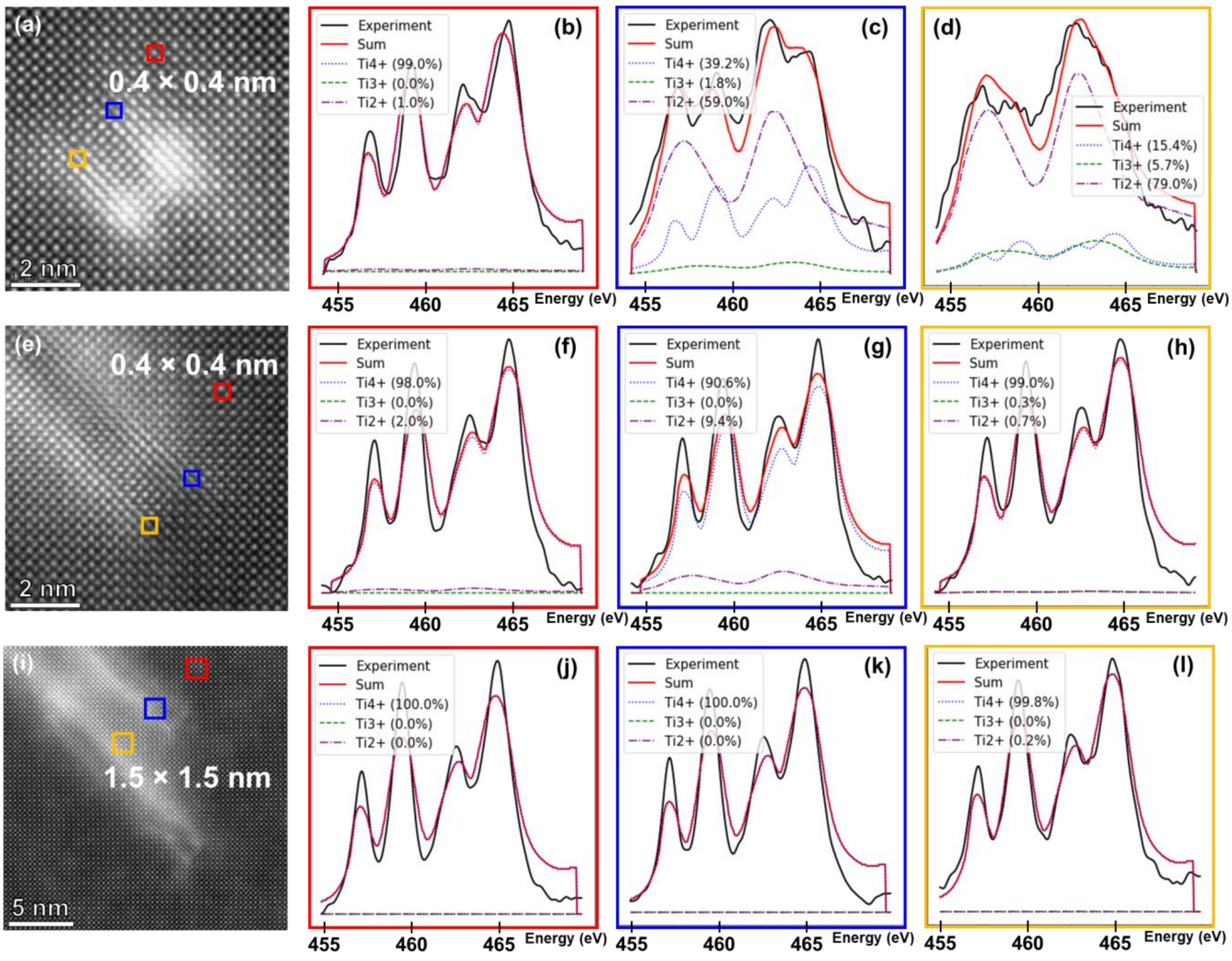

**Figure 2. High resolution MAADF image and corresponding EELS spectra on ⟨110⟩{$\bar{1}$10} dislocation loop.** (a, e, i) high resolution MAADF image of (a) edge core of relatively short loop, (e) edge core of relatively long loop and (i) screw component of relatively long loop, respectively. (b- d) EELS spectra and NNLS results on each (b) red, (c) yellow, (d) blue square on (a). $Ti^{2+}$ to $^{4+}$ was used as component and reference peaks of them were taken from previous research [29,30], and percentages in the parentheses are weight of each component. (f-h) EELS spectra and NNLS results on each (f) red, (g) yellow, (h) blue square on (e). (j-l) EELS spectra and NNLS results on each (j) red, (k) yellow, (l) blue square on (i).

Spectroscopy investigations were performed from the edge cores of a near-point-like dislocation loop with a diameter of several nanometers (**Figure 2a-d**), the edge core and the screw segment region (**Figure 2i-l**) (at the end of a larger dislocation loop with a diameter of tens of nanometers (**Figure 2e-h**), respectively. In addition to the screw-dominant segment identified in the low-magnification BF-TEM image, edge cores located at the loop termination were also directly observed in the atomic-resolution STEM images. EELS spectra were acquired in red, yellow, and blue square-indicated regions in the MAADF STEM images. We analyze these spectra using non-negative least square (NNLS) fitting with reference EELS spectra with known oxidation states ($Ti^{4+}$ from $SrTiO_3$, $Ti^{3+}$ from $Ti_2O_3$, and $Ti^{2+}$ from TiO; Details about EELS process are described in Supplementary Materials **S1**) [29,30]. The EELS spectrum taken near the edge component of a dislocation core in the smaller loop in **Figure 2a** is illustrated in **Figure 2b-d**. In the red square, which is far from the edge-dislocation core, the Ti L-edge peaks mainly consist of $Ti^{4+}$. However, in the blue and yellow squares, which are located at the edge dislocation core (in **Figure 2c,d**), the NNLS results reveal not only $Ti^{4+}$ but also $Ti^{2+}$ and $Ti^{3+}$ peaks. Notably, the spectrum in the yellow square shows a higher $Ti^{2+}$ weight fraction (75.7%) than that from the blue square (53.8%). This is reasonable, as a full ⟨110⟩ dislocation consists of both Sr and Ti planes. When it decomposes, one dislocation core will be Ti-rich, while the other will be Sr-rich. Only the Ti-rich core will exhibit a clear change in the Ti oxidation state. For the edge core on the larger dislocation loop in **Figures 2e-h**, the Ti L edge peaks in a red square (far from the edge dislocation core) resemble those in pristine $SrTiO_3$ [29,30]. However, the yellow and blue squares, which correspond to the edge dislocation core, are dominated by $Ti^{4+}$ and are similar to those of the pristine lattice. Similarly, for screw component on the dislocation loop, the Ti L-edge peaks from the regime in red, yellow, and blue squares are all dominated by $Ti^{4+}$, as in **Figure 2i,l**. The region analyzed in **Figure 2i-l** was selected from the dislocation segment identified as screw-dominant from the low-magnification BF-TEM two-beam condition analysis described above **(Figure 1e,f)**. The small mismatch between the reconstructed spectrum (NNLS sum) and the measured spectrum mainly induced by differences in energy resolution between our dataset and the literature reference spectra. We repeated the same STEM–EELS acquisition and NNLS-based Ti-valence analysis on additional dislocation loops (two more short and long loops each), obtaining consistent oxygen-deficiency signatures for short loops and near-stoichiometric signatures for long loops (Supplementary Materials: Figure S2).

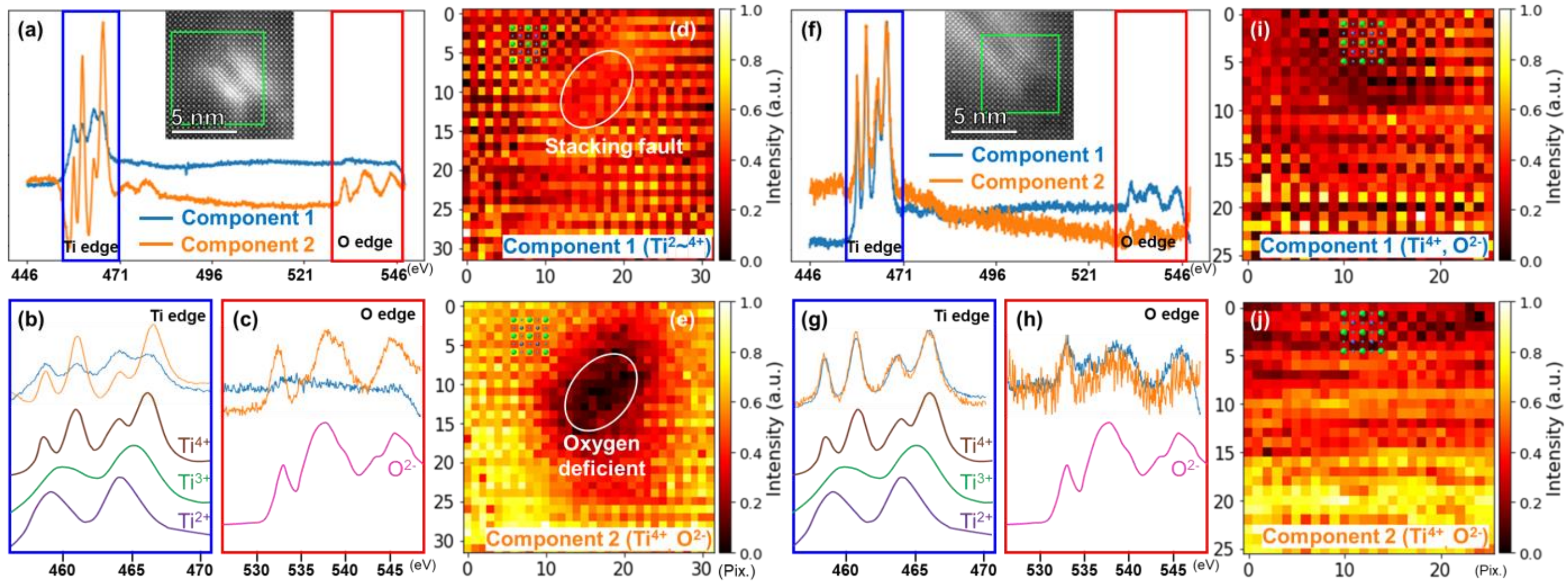

**Figure 3. PCA result of EELS spectrum on relatively short loop edge core and relatively long loop.** (a) PCA decomposition of small dislocation loop; the inset shows the corresponding STEM image from Figure 2a, and the EELS mapping region used for the PCA analysis is marked by a green square. (b) Comparison of Ti L-edge peaks from PCA results with standard reference peaks for $Ti^{2+}$–$Ti^{4+}$ ions. (c) Comparison of O K-edge peaks from PCA results with the standard $O^{2-}$ peak in $SrTiO_3$. (d) Spatial distribution of component 1 which is related with $Ti^{2+}$ to $Ti^{4+}$ ion. (e) Spatial distribution of component 2 which is related with $Ti^{4+}$ and $O^{2-}$ ion. (f) PCA decomposition near the edge core of the long dislocation loop; the inset shows the corresponding STEM image from Figure 2e, and the analyzed EELS mapping region is marked by a green square. (g) Comparison of Ti L-edge peaks from PCA results with standard reference peaks for $Ti^{2+}$–$Ti^{4+}$ ions. (h) Comparison of O K-edge peaks from PCA results with the standard $O^{2-}$ peak in $SrTiO_3$. (i, j) Spatial distribution of (i) component 1 and (j) component 2.

Principal component analysis (PCA) was performed on the EELS spectrum images to further investigate the gliding of dislocation core during plastic deformation. (Details about EELS process are described in Supplementary Materials **S1**). **Figure 3a** shows the decomposing EELS spectrum image from **Figure 2a** into two components with PCA. The corresponding STEM image from **Figure 2a** is included as an inset of **Figure 3a**, and the EELS mapping region used for the PCA analysis is marked by a green square. We compared these two components at the Ti L-edge energy range (456–471 eV) against several reference peaks with known Ti oxidation states, as presented in **Figure 3b** [29,30]. Component 1 (blue curve) contains contributions not only from the $Ti^{4+}$ peak of reference $SrTiO_3$, but also from $Ti^{3+}$ ($Ti_2O_3$) and $Ti^{2+}$ (TiO) [29]. In contrast, Component 2 (orange curve) matches the $Ti^{4+}$ peak of $SrTiO_3$ [30]. Additionally, we analyzed each component's O K-edge range (525–550 eV) and compared them to the $O^{2-}$ reference spectrum (**Figure 3c**). Component 1 exhibited no distinct oxygen peak, whereas Component 2 is similar to the oxygen peak of single-crystal $SrTiO_3$ [30].

Based on the component analysis, we generated relative spatial distribution maps in arbitrary units of Component 1 (representing spectral features resembling reduced-Ti-like ELNES variation) and Component 2 (resembles stoichiometric STO-like spectral variation), shown in **Figure 3d**,**e**, respectively. (NNLS decomposition of the Ti $L_{2,3}$ edge map into $Ti^{2+/}Ti^{3+}/Ti^{4+}$ components reproduces the same spatial trend of PCA, Supplementary Materials: **Figure S3**) Each pixel in the EELS spectrum image corresponds to approximately 0.2 nm, which is about half the $SrTiO_3$ lattice parameter (~0.4 nm). Therefore, the pixelated square pattern in Figure 3d,e reflects the discrete STEM-EELS scan grid used for spectral acquisition and visualization, rather than a physical square-shaped structural feature. Given that the spatial resolution of STEM-EELS approaches that of the electron probe size [42,43], the relatively bright regions in **Figure 3d** are primarily attributed to the Ti in Component 1. The analyzed spectrum range (446 to 548.4 eV) excludes Sr signals (L-edge at 1940–2220 eV, M-edge at 133–358 eV), so Sr sites appear as dark spots. In the map of Component 2, which resembles the stoichiometric STO-like $Ti^{4+}$ and O-K fine structure, the regions around the edge core and stacking fault appear darker [42,43], consistent with locally reduced O-K spectral weight and oxygen-vacancy in these regions. In other words, oxygen vacancy signatures tend to localize near the edge core in case of relatively short dislocation.

**Figure 3f** is also the PCA result of EELS spectrum image of **Figure 2e**, which is from a dislocation with further gliding (the analyzed EELS mapping region indicated as green square in the inset image). The decomposed components appeared almost identical in shape, as shown in **Figure 3g,h**, indicating that the Ti oxidation state remains unchanged from pristine STO. **Figure 3i,j** presents the relative spatial distributions maps in arbitrary units of Components 1 and 2, respectively, for the longer dislocation. No obvious oxygen-deficient region is observed since both components exhibit nearly identical spectral shapes.

To corroborate with the experimental observation, we further performed the computational analysis on: (i) the average oxygen deficiency state around a dislocation, and (ii) the effective charges on the Ti atoms at different distances away from the dislocation core through atomic-level MD simulations. **Figure S4** presents the set-up of our MD model where a dislocation of edge character and [110] Burgers vector was initially introduced in the $(\bar{1}10)$ plane and then relaxed to dissociate into two partials with a stacking fault in between. In order to quantify the defect chemistry change charge state change during its motion, similar to that in **Figure S4**, an oxygen deficient (model "positively charged") dislocation core was built into the sample. The equilibrated model containing such a model-charged dislocation was then deformed under shear. The applied shear, noted as $\tau_{ap}$, was increased from 0 to 3 GPa, then held as constant at 3 GPa. Such a high stress is imposed here because it facilitate and speed up the motion of a dislocation in $SrTiO_3$ because the minimum stress required to enable the motion of it in current MD models was found to be ~1.5 GPa. This is significantly higher than the Peierls stress for $SrTiO_3$, but is commonly used in nanoscale MD simulations. Another reason for the deployment of such a high stress is that only a very short dislocation segment is considered in a limited simulation cell without considering any pre-existing defects. As shown in [44], if needed, the stress required to facilitate the motion of a dislocation in high-Peierls-barrier materials can be largely reduced when a micrometer-long dislocation is considered by allowing collective kink activities on it. Results from MD simulations at a relatively lower stress, 1.5 GPa, are found to be similar as that at 3.0 GPa and included in the supplemental materials for comparison.

To characterize the local defect-chemical state, two strategies have been deployed. One is for the overall charge distribution using a Gaussian approach, in which the Gaussian-smoothed field of assigned ionic charges at any point with a coordinate of $(x, y, z)$ in the simulation cell can be calculated according to the contributions from the charges on the atoms surrounding that point. Such calculation will give rise to a charge distribution field, which evolves when dislocation moves. The other is for calculating the effective charges on the Ti atoms, in which the number of O atoms surrounding each Ti atom in both reference and current configuration is counted and then used to assess the effective charges on Ti atoms (see more details in Supplementary Materials Part 2 for MD model set-up, boundary conditions, interatomic potential, and the defect chemistry characterization strategy).

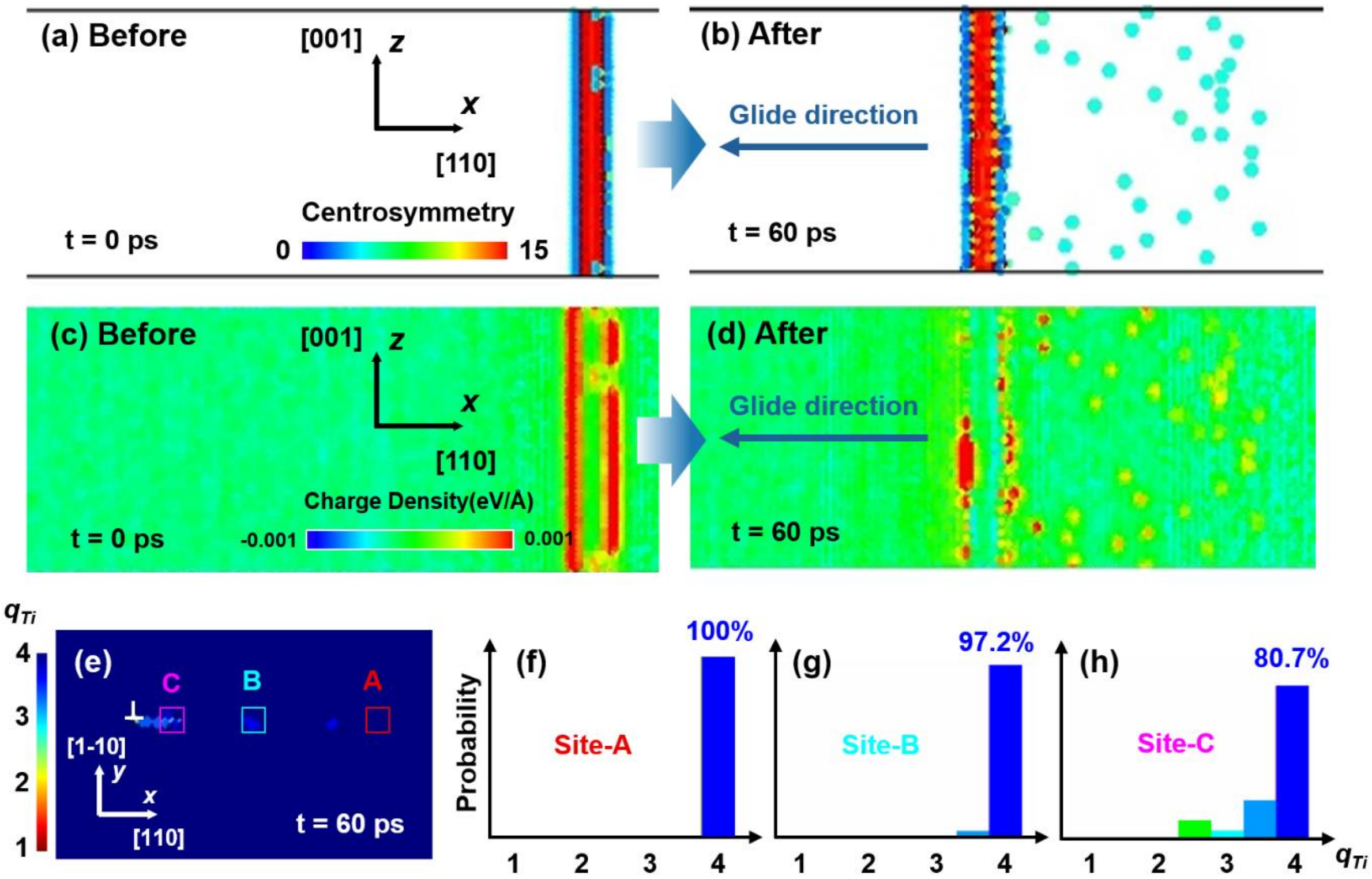

**Figure 4. MD simulation results for a single-crystalline $SrTiO_3$ sample containing a ⟨110⟩{$\bar{1}$10} edge dislocation under a shear stress of 3.0 GPa.** (a, b) Time-sequence snapshots of the dislocation motion, showing dislocations (blue), stacking faults (red), and oxygen vacancies (light blue). Atoms are color-coded using the centrosymmetry parameter in OVITO [45], and only those involved in the defects are displayed. (c, d) Evolution of the charge density distribution (model proxy) during dislocation motion; red, blue, and green correspond to positive, negative, and neutral charge states, respectively. Panels (a, c) represent the initial state (t = 0 ps), while (b, d) correspond to the state after motion (t = 60 ps). (e–h) Computational analysis of the effective charge on Ti atoms ($q_{\mathrm{Ti}}$). (e) Positions of the three representative

sites selected for analysis. (f) Site-A (far from the dislocation core). (g) Site-B (the region around one vacancy). (h) Site-C (nearby the dislocation core).

**Figure 4** presents time sequences of the dislocation motion (**Figure 4a,b**) and the corresponding local charge state evolution (**Figure 4c,d**) resulting from the MD simulations at 3.0 GPa. Three main observations are: (i) the dislocation lines in $SrTiO_3$ do not move as a whole; similar to many other materials with high Peierls barrier [46], the motion occurs through kink activation and migration along the dislocation line; (ii) this rough, kink-mediated motion leads to the production of point defects (light blue in **Figure 4a,b**) on the slip plane where the dislocation belongs to. A local atomic structure analysis confirms these point defects to be oxygen vacancies; and (iii) an initially oxygen-deficient dislocation (red in **Figure 4c** when $t$ = 0 ps) becomes less oxygen-deficient and approaches a near-stoichiometric signature as it moves rapidly under 3 GPa (green in **Figure 4d** when $t$ = 60 ps). Accompanied by such charge state change, oxygen vacancies are formed and left behind the moving dislocation, implying that the initially oxygen-deficient core condition evolves during glide while leaving an oxygen-deficient trail.

As further evidence of the dislocation motion-induced local charge change, consistent with experimental observations, the effective charges on Ti atoms, noted as $q_{Ti}$, at three representative sites (i.e., far away, near, and at the dislocation core) were calculated and analyzed (**Figure 4e-h).** The population of Ti atoms with an effective charge of +1, +2, +3, and +4 at these sites (A, B, and C) was counted. Results show that the majority of Ti atoms carry an effective charge of +4 (blue, noted as $Ti^{+4}$), while the deviations from +4 occurs near or at the core. The percentage of $Ti^{+4}$ at the dislocation core (80.7% at Site-C) is much lower than that (100% for Site-A) away from the core. Moreover, the local atomic structure differences cause the charge state in Site-B (the region around one vacancy) to differ from that in Site-A. According to the definition of the effective charges on Ti atoms given in Supplemental Materials, the charge on Ti deviates from +4 when the population of its oxygen neighbor, noted as $n_O$, differs from that in a reference configuration, $n_{O\text{-}ref}$. In particular, for $Ti^{+1}$, $Ti^{+2}$, and $Ti^{+3}$ atoms, $n_O$ is smaller than $n_{O\text{-}ref}$, confirming that kink-induced oxygen-vacancy production contributes directly to the change of the local charge states.

On the other hand, during dislocation motion, a near-stoichiometric core region may become more oxygen-deficient through interaction with oxygen vacancies. However, because the oxygen vacancy density is much lower than that of perfect lattice fraction, this process is far less likely than an initially oxygen-deficient core condition relaxing while leaving oxygen vacancies behind.

These MD results indicate that an initially oxygen-deficient dislocation core can evolve toward a less oxygen-deficient state during kink-mediated glide, while leaving oxygen vacancies behind on the slip plane. As the dislocation glides and its line length increases, maintaining such an oxygen-deficient configuration becomes increasingly unfavorable; consequently, vacancy emission and progressive charge neutralization are energetically preferred. The dislocation migrates as a whole via lower-height kinks compared to the initial state, thereby achieving and maintaining overall charge neutrality. This trend is qualitatively consistent with the experimentally observed contrast between short loops, which exhibit stronger oxygen-deficient signatures near the edge core, and longer loops, which remain closer to stoichiometry. Additional analyses of the charge-state evolution and dislocation-line configuration during glide are provided in the Supplementary Materials **Figure S6**.

Overall, the MD simulation results are qualitatively consistent with experimental observations for capturing the essential physics. We, however, also note the limitations of the modeling and its qualitative description at this stage due to its two main limitations: (i) a rigid-ion interatomic potential has been used here and thus explicit charge transfer and redox are not considered. An inclusion of the charge transfer into the interatomic potential is needed if one desires to quantitively capture the electronic-structure evolution associated with oxygen non-stoichiometry during the process of dislocation motion; (ii) with a nano-sized simulation cell size, the current MD model can only accommodate nanometer-long dislocations, which is far less than the dislocation line length in experiments. To fill such a length scale gap, the mesoscale simulation tool, such as the coarse-grained atomistic models in [44,47], that can accommodate the micrometer-long dislocation, nanoscale kink dynamics, and atomic-scale vacancy all in one model will be necessary. Such simulations endeavor opens new challenges but will eventually get one step closer to that in experiments.

Last but not least, contrasting the key role that defect chemistry plays for dislocations in ionic crystals, the dislocation study in ceramics has so far been mostly referred to those in metals for analogy or to circumvent the complexity. This has been rather successful in dislocation engineering and mechanical deformation across the length scales, but this metal/ceramics dislocation analogy approach may have overlooked the key differences [48]. It remains a challenging but interesting puzzle on the fundamental mechanisms for the dozens of ceramics exhibiting room-temperature bulk plasticity mediated by dislocations [49]. The oxygen vacancy of the dislocation cores in $SrTiO_3$ is demonstrated here to effectively impact the dislocation mobility. This highlights one of these key differences between ceramics and metals, providing a useful perspective for understanding dislocation plasticity in oxide ceramics.

## 4. Conclusions

To summarize, by integrating high-resolution imaging and spectroscopy with atomistic simulation, we characterize the local chemistry of dislocations in plastically deformed $SrTiO_3$. STEM–EELS shows that small dislocation loops exhibit an obvious Ti reduction and oxygen vacancy enrichment at their edge core, whereas large loops remain dominated by $Ti^{4+}$ and are effectively neutral. Atomic-level MD simulations capture the same mechanistic picture: kink-assisted-dislocation glide leaves an oxygen-vacancy-rich trail and drives transition from nonstoichiometric to stoichiometric. These convergent observations revealed oxygen-vacancy-dependent dislocation mobility in plastically deformed $SrTiO_3$ where stoichiometric dislocations glide most readily. Our work explains why morphologically similar dislocations show disparate glide distances and highlights defect control, via vacancy or local chemistry management, as a strategy to tailor both plasticity and functionality in functional oxides.

## Acknowledgements

MC and LZ acknowledge the support faculty startup funding. YC and LX acknowledge the support of the US National Science Foundation (CMMI-2322675 and CMMI-2328533). XF and AF acknowledge the European Research Council (ERC) for funding support (ERC Starting Grant, grant number 101076167, project MECERDIS). LX also acknowledges the supplemental funds from US National Science Foundation for supporting his collaboration with the ERC-funded group led by XF.